\documentstyle[a4,11pt]{article}

\begin{document}
\hfill
NIKHEF-H/92-18

\begin{center}
{\huge The renormalization of the axial anomaly \\
 in dimensional regularization} \\
\vspace{1cm}
{\bf S.A. Larin}
\footnote{on leave from Institute for Nuclear Research (INR)
of the Russian Academy of Sciences, Moscow 117312, Russia.} \\
\vspace{2mm}
NIKHEF-H \\ P.O. Box 41882 \\ 1009 DB Amsterdam. \\ \vspace{5mm}
\end{center}

\begin{abstract}
The prescription for the $\gamma_5$-matrix
within dimensional regularization in multiloop calculations is elaborated.
The three-loop anomalous dimension of the singlet axial current is
calculated.
\end{abstract}

\newpage
Since the invention of dimensional regularization \cite{hv} and the
minimal subtraction ($MS$) scheme \cite{h}
a lot of attention was paid to the problem of the $\gamma_5$-matrix
within dimensional regularization.
The following approaches to this problem were used in practical
calculations:
the prescriptions based on the original definition by 't Hooft and Veltman
\cite{hv} \cite{ad} \cite{bm}, keeping the four-dimensional
anticommutation relation for $\gamma_5$ in $D$-dimensions \cite{cfh} and
dimensional reduction \cite{s}.
Discussions of the $\gamma_5$-prescriptions
can be found e.g. in \cite{collins} \cite{buras}.

The effective approach to perform multiloop calculations involving the
non-singlet axial current in dimensional regularization was developed
in \cite{gl} \cite{lv} where deep inelastic sum rules were calculated
up to (and including) the three-loop level in QCD.
The effectiveness of this approach is confirmed by its use in such an
advanced calculation as the calculation of the deep
inelastic structure function $F_3$ at the two-loop QCD level \cite{zn}.

In the present paper this approach is elaborated for the
cases of the pseudoscalar current and the singlet axial current.
The three-loop anomalous dimension
of the singlet axial current is calculated by imposing
the requirement that the axial anomaly relation \cite{adler} \cite{bj}
should preserve the one-loop character \cite{ab} in
dimensional regularization.

Throughout the paper we use the $MS$-scheme
\cite{h} or its standard modification, $\overline{MS}$-scheme \cite{msb},
to perform renormalizations. The dimension of space-time
is defined in the standard way as $D=4-2\epsilon$.
All calculations are performed within massless perturbative QCD.

{\bf 1. The non-singlet axial current.}
Let us first consider the case of the non-singlet axial current:
\begin{eqnarray}
\label{ans}
J^{5a}_\mu(x) = \overline\psi(x) \gamma_\mu \gamma_5 t^a \psi(x),
\end{eqnarray}
where $\psi$ is a quark field and $t^a$ is a generator of a flavor
group.

In our opinion, the most practical definition of $\gamma_5$ for
multiloop calculations in dimensional regularization
(and the only one known to be self-consistent) is
the original definition due to 't Hooft and Veltman \cite{hv}:
\begin{eqnarray}
\label{g5}
	\gamma_5  = i \frac{1}{4!} \varepsilon_{\nu_1\nu_2\nu_3\nu_4}
	\gamma_{\nu_1} \gamma_{\nu_2} \gamma_{\nu_3} \gamma_{\nu_4},
\end{eqnarray}
here the Levi-Civita $\varepsilon$-tensor is unavoidably a four-dimensional
object and
should be taken outside the $R$-operation where any object can be safely
considered as a four-dimensional one;
the indices $\nu_1 \ldots \nu_4$ are $D$-dimensional inside the $R$-operation
as all other indices within dimensional regularization.
But $\gamma_5$ defined by eq.(\ref{g5}) does not anticommute anymore
with the $D$-dimensional $\gamma_{\mu}$. That is why in order
to define the axial
current correctly one should use (see below comments after eq.(\ref{z5}))
the symmetrical form of the axial current:
\begin{eqnarray}
\label{sans}
J^{5a}_\mu = \frac{1}{2} \overline\psi (\gamma_\mu \gamma_5
-\gamma_5\gamma_\mu) t^a \psi,~~~~
\gamma_5  = i \frac{1}{4!} \varepsilon_{\nu_1\nu_2\nu_3\nu_4}
	\gamma_{\nu_1} \gamma_{\nu_2} \gamma_{\nu_3} \gamma_{\nu_4}.
\end{eqnarray}
In principle it is possible to perform the calculations
using this definition of the axial current.
But one can simplify the definition drastically. Let us commute
$\gamma_\mu$ in the first term in (\ref{sans}) to the right.
The $D$-dimensional metric tensors
$g_{\mu\nu_i}$ arising during commutations can always be taken outside the
$R$-operation where they can be safely contracted with the
$\varepsilon$-tensor as four-dimensional objects. So we receive
the following definition of the non-singlet axial current:
\begin{equation}
\label{adc}
J^{5a}_\mu =i \frac{1}{3!}\varepsilon_{\mu\nu_1\nu_2\nu_3}
	\overline\psi \gamma_{\nu_1} \gamma_{\nu_2} \gamma_{\nu_3} t^a \psi.
\end{equation}
This is exactly the prescription proposed in \cite{ad}. Thus we proved
equivalence of the definitions (\ref{adc}) and (\ref{sans}) within dimensional
regularization. To be sure that no holes
are missed in this general proof we computed the axial vertex
(see below eq.(\ref{recep})) at the three-loop level, using
both definitions. The results are identical. But the definition
(\ref{adc}) is more compact and saves computational time enormously so we
will use it in what follows.

Since the anticommutativity of $\gamma_5$ is violated by definition
(\ref{g5}), the standard properties of the axial current and Ward identities
which are valid e.g.
in such a basic regularization as the Pauli-Villars regularization are also
violated.
In particular the renormalization constant $Z_{MS}^{ns}$ of the non-singlet
axial
current in the $MS$-scheme is not equal to one any more. It was calculated
in the three-loop approximation in QCD in \cite{lv}; we remind here the
two-loop expression:
\begin{eqnarray}
\label{za}
Z_{MS}^{ns} & = &
      1 + a^2 \frac{1}{\epsilon} ( \frac{22}{3}C_FC_A - \frac{4}{3}C_Fn_f ),
\end{eqnarray}
where we use the notation $a=\frac{\alpha_s}{4\pi}=\frac{g^2}{16\pi^2}$
for the strong coupling constant,
$C_F$ and $C_A$ are the Casimir operators of the defining and the
adjoint representations of the color group and $n_f$ is the number of quarks
flavors. The relation between renormalized and bare operators is
$(O)_R=Z~(O)_B$.

To restore the renormalization invariance of the non-singlet axial current
(i.e. to nullify its anomalous dimension)
one should perform \cite{t} an extra
finite renormalization or in other words to introduce
the extra finite renormalization constant $Z_5^{ns}(a)$.
So the correct renormalized non-singlet axial current is:
\begin{eqnarray}
\label{phys}
(J^{5a}_\mu)_R=Z_5^{ns}(a)~Z_{MS}^{ns}(a)~ (J^{5a}_\mu)_B=
Z_5^{ns}~Z_{MS}^{ns}~
       i \frac{1}{3!}\varepsilon_{\mu\nu_1\nu_2\nu_3}
\overline\psi_B \gamma_{\nu_1} \gamma_{\nu_2} \gamma_{\nu_3} t^a \psi_B,
\end{eqnarray}
where $\psi_B=Z_2^{\frac{1}{2}}\psi$ is a bare quark field.
The full anomalous dimension can be now nullified:
\begin{eqnarray}
\label{zero}
\gamma_J^{ns}(a)=\mu^2\frac{d}{d\mu^2}\log(Z_5^{ns}~Z_{MS}^{ns})=
\beta (a)\frac{\partial \log Z_5^{ns}}{\partial a}
+\mu^2\frac{d}{d\mu^2}\log Z_{MS}^{ns} =0.
\end{eqnarray}
Using this equation one can obtain $Z_5^{ns}$ from the given $Z_{MS}^{ns}$.
But since the renormalization group $\beta$-function starts with an $a^2$-term
one can obtain $Z_5^{ns}(a)$ only in the approximation which is one order
in $a$ less than the given approximation of $Z_{MS}^{ns}$.
That is why it is better to use the recipe of \cite{gl} and to find $Z_5^{ns}$
from the requirement that the renormalized axial and vector
vertices coincide:
\begin{eqnarray}
\label{recep}
Z_5^{ns} R_{MS} <\overline{\psi}~ J^{5a}_\mu(0)~ \psi>=
R_{MS} <\overline{\psi}~ J_\mu^a(0)~ \psi >\gamma_5,
\end{eqnarray}
where $J_\mu^a(x)=\overline\psi(x) \gamma_\mu \gamma_5 t^a \psi(x)$ is the
vector current.
This relation means that anticommutativity of the $\gamma_5$-matrix is
effectively restored, so the standard Ward identities are also restored.
This prescription for $Z_5^{ns}$ automatically ensures
zero anomalous dimension (\ref{zero}) because the anomalous dimension
of the vector vertex is naturally zero.

The first impression is that the calculation of the axial vertex is rather
cumbersome within adopted prescription (\ref{adc}) for the axial current:
taking
the $\varepsilon$-tensor outside $R$-operation seems to leave inside
$R$-operation
three extra uncontracted indices $\nu_1,\nu_2,\nu_3$.
But it is always possible
to contract first a quantity under consideration with an extra
$\varepsilon$-tensor to produce a scalar object.
In our case we first
multiply the axial vertex with $\gamma_\mu\gamma_5=i
\frac{1}{3!}\varepsilon_{\mu\rho_1\rho_2\rho_3}
	\gamma_{\rho_1} \gamma_{\rho_2} \gamma_{\rho_3}$
and take the trace. Now outside the $R_{MS}$-operation the product of
four-dimensional $\varepsilon$-tensors
can be represented in a standard way as the determinant of the
four-dimensional
metric tensors:
\begin{eqnarray}
\label{eps}
\varepsilon_{\mu\nu_1\nu_2\nu_3}\varepsilon_{\mu\rho_1\rho_2\rho_3}=
g_{\nu_1\rho_1}(g_{\nu_2\rho_2}g_{\nu_3\rho_3}-
                 g_{\nu_2\rho_3}g_{\nu_3\rho_2})+...
\end{eqnarray}
Then outside the $R_{MS}$-operation these
four-dimensional metric tensors can be safely considered as $D$-dimensional
ones (which will add only inessential $O(\epsilon)$-terms to
the renormalized axial vertex). These $D$-dimensional metric tensors
$g_{\mu_i\rho_j}$ can be safely taken inside the $R$-operation. So one obtains
finally a scalar expression inside the $R$-operation containing only
$D$-dimensional
objects which makes the practical calculations straightforward.

Calculating the axial vertex and
the vector vertex in eq.(\ref{recep}) one finds $Z_5^{ns}$.
The three-loop approximation for $Z_5^{ns}$ was obtained in \cite{lv}; we
recall here the two-loop expression:
\begin{eqnarray}
\label{z5}
Z_5^{ns}  =  1
       + a  (  - 4C_F )
       + a^2  (  22C_F^2 - \frac{107}{9}C_FC_A + \frac{2}{9}C_Fn_f  ).
\end{eqnarray}
The independence of $Z_5^{ns}$ on the $\log(p^2/\mu^2)$ (where $p$ is the
momentum
of quark legs in the vertices in eq.(\ref{recep})) gives a strong check of the
whole prescription. For example, trying to use the axial current in the
naive form
(\ref{ans}) with $\gamma_5$ defined in (\ref{g5}) one would obtain that
$\log(p^2/\mu^2)$ does not cancel in $Z_5^{ns}$ which excludes the possibility
to use this naive form.

Note that both $Z_{MS}^{ns}$ and $Z_5^{ns}$ are gauge independent quantities
(which was checked by calculations in an arbitrary covariant gauge).
We would like to note also that the finite constant $Z_5^{ns}(a)$, like
the usual ultraviolet renormalization
constants, does not depend on the choice of the concrete modification of the
$MS$-like schemes: whether it is calculated within the $MS$-scheme itself or
$\overline{MS}$-scheme or G-scheme \cite{G}.

So to calculate any quantity involving the non-singlet axial current
one can use the prescription (\ref{phys}) with $Z_{MS}^{ns}$ and $Z_5^{ns}$
given in (\ref{za}) and (\ref{z5}). This prescription has all typical features,
see \cite{collins}, of the
approach developed in \cite{bm}, where $D$-dimensional indices split
into 4-dimensional and $(D-4)$-dimensional indices.
But in our approach all indices during the calculations are
$D$-dimensional which avoids in practical calculations
all complications connected with splitting indices.

In the case of several $\gamma_5$-matrices
in one fermion line one can naively anticommute them
(because of the validity of eq.(\ref{recep}))
as in the prescription
with anticommuting $\gamma_5$ \cite{cfh} and use the standard
property $\gamma_5^2=1$. So e.g. the correlator of two axial non-singlet
currents automatically coincides in the approach under consideration with
the correlator of two vector non-singlet currents.

{\bf 2. The pseudoscalar current.}
Let us apply now this prescription to the case of the pseudoscalar
current:
\begin{eqnarray}
\label{p}
P(x)=\overline\psi(x) \gamma_5 \psi(x)=
i\frac{1}{4!} \varepsilon_{\nu_1\nu_2\nu_3\nu_4}
\overline\psi \gamma_{\nu_1} \gamma_{\nu_2} \gamma_{\nu_3} \gamma_{\nu_4} \psi
\end{eqnarray}
We will not distinguish singlet and non-singlet cases for the pseudoscalar
current because the
pseudoscalar current does not generate closed fermion loops in the massless
case.
As in the case of the non-singlet axial current, to obtain the correct
renormalized pseudoscalar current we should introduce the finite constant
$Z_5^p$ in addition to the usual ultraviolet renormalization
constant $Z_{MS}^p$
in the MS-scheme:
\begin{eqnarray}
\label{pr}
(P)_R=Z_5^p(a)Z_{MS}^p(a)(P)_B=Z_5^pZ_{MS}^p
i\frac{1}{4!} \varepsilon_{\nu_1\nu_2\nu_3\nu_4}
\overline\psi_B \gamma_{\nu_1} \gamma_{\nu_2} \gamma_{\nu_3} \gamma_{\nu_4}
\psi_B
\end{eqnarray}
We can calculate first in the standard way the renormalization constant
$Z_{MS}^p$ within the MS-scheme.
In all calculations we use the program Mincer \cite{mincer} written for
the symbolic manipulation system Form \cite{form}. This program computes
analytically the three-loop massless diagrams of propagator type which
is sufficient to calculate any renormalization constant within MS-scheme
at the three-loop level.

The three-loop approximation for $Z_{MS}^p$ in the MS-scheme is:
\begin{eqnarray}
\label{zp}
Z_{MS}^p &=& 1+ a(  -C_F \frac{3}{\epsilon} )
       + a^2[C_FC_A (\frac{11}{2\epsilon^2} + \frac{79}{12\epsilon} )
       + C_Fn_f(  -\frac{1}{\epsilon^2} - \frac{11}{6\epsilon} )
       + C_F^2 (\frac{9}{2\epsilon^2} - \frac{3}{4\epsilon}) ]
\nonumber \\ &&
       + a^3[C_FC_An_f( \frac{44}{9\epsilon^3} + \frac{110}{27\epsilon^2}
 +\frac{8\zeta_3}{\epsilon} - \frac{58}{9\epsilon} )
       + C_FC_A^2 ( - \frac{121}{9\epsilon^3} - \frac{257}{54\epsilon^2}
 -\frac{599}{108\epsilon} )
\nonumber \\ &&
       + C_Fn_f^2( - \frac{4}{9\epsilon^3} - \frac{22}{27\epsilon^2}
 + \frac{17}{27\epsilon} )
       + C_F^2C_A ( - \frac{33}{2\epsilon^3} - \frac{215}{12\epsilon^2}
 + \frac{3203}{36\epsilon} )
\nonumber \\ &&
       + C_F^2n_f ( \frac{3}{\epsilon^3} + \frac{19}{6\epsilon^2}
 - \frac{8\zeta_3}{\epsilon} - \frac{107}{9\epsilon} )
       + C_F^3 ( - \frac{9}{2\epsilon^3} + \frac{9}{4\epsilon^2}
 - \frac{43}{2\epsilon} )],
\end{eqnarray}
where $\zeta_3$ is the Riemann zeta-function ($\zeta_3 =
1.202056903\ldots$).
To restore the Ward identities we can, as in the
case of the non-singlet axial current,
define the finite renormalization constant $Z_5^p$
from the requirement of coincidence of the pseudoscalar and scalar vertices:
\begin{eqnarray}
\label{recep2}
Z_5^p R_{MS} <\overline{\psi}~ P(0)~ \psi>=
R_{MS} <\overline{\psi}~~ \overline\psi \psi(0)~~ \psi>\gamma_5,
\end{eqnarray}
So the
anticommutativity of the  $\gamma_5$-matrix is effectively restored
and it is anticommuted out of the pseudoscalar vertex.
Calculating the three-loop pseudoscalar and scalar vertices we find the
three-loop approximation for $Z_5^p$:
\begin{eqnarray}
\label{z5p}
Z_5^p &=& 1 + a  (  - 8C_F )+ a^2  ( \frac{2}{9}C_FC_A + \frac{4}{9}C_Fn_f )
 +a^3  [ C_FC_An_f(\frac{64}{3}\zeta_3
+ \frac{856}{81})
\nonumber \\ &&
+C_FC_A^2(- 208\zeta_3 - \frac{958}{27})
+ \frac{104}{81}C_Fn_f^2 +C_F^2C_A( 608\zeta_3
- \frac{800}{27})
\nonumber \\ &&
+C_F^2n_f( - \frac{64}{3}\zeta_3
- \frac{580}{27}) +C_F^3(- 384\zeta_3 + \frac{304}{3} ) ]
\end{eqnarray}
Again the cancellation of $\log(p^2/\mu^2)$ in $Z_5^p$ provides a good
check of the calculations. Another check is the gauge independence of
both $Z_{MS}^p$ and $Z_5^p$ (all calculations were done in an arbitrary
covariant
gauge). Note that $Z_5^p$ for the pseudoscalar current differs from
$Z_5^{ns}$ for the non-singlet axial current.

Note also that the full renormalization constant $Z_5^p Z_{MS}^p$
of the pseudoscalar current
does not coincide with the renormalization constant of the
scalar current $Z_{\overline\psi \psi}$ but their anomalous dimensions do
coincide:
\begin{eqnarray}
\label{coin}
\mu^2\frac{d}{d\mu^2}\log (Z_5^p~Z_{MS}^p)=
\mu^2\frac{d}{d\mu^2}\log Z_{\overline\psi \psi}=
       + a ( 3C_F )
       + a^2  ( \frac{3}{2}C_F^2 +\frac{97}{6}C_FC_A - \frac{5}{3}C_Fn_f  )
\nonumber \\
 +a^3~[~\frac{129}{2}C_F^3 -\frac{129}{4}C_F^2C_A +\frac{11413}{108}C_FC_A^2
\nonumber \\
+C_F^2n_f (24\zeta_3 - 23 )
+C_FC_An_f( - 24\zeta_3 - \frac{278}{27})
 - \frac{35}{27}C_Fn_f^2 ].
\end{eqnarray}
We would like to note that this our result agrees with the original
calculation of the three-loop anomalous
dimension of the quark mass in the $MS$-scheme in \cite{tar} and provides
thus the independent check of that calculation.

Thus one can use in all calculations involving the pseudoscalar current
the prescription (\ref{pr}) with
$Z_5^p$ and $Z_{MS}^p$ given in (\ref{z5p}) and (\ref{zp})
at the three-loop level.
In the case of several pseudoscalar vertices
in one fermion line one can
(since the anticommutativity of the $\gamma_5$ is effectively restored
by the prescription (\ref{recep2}))
naively anticommute $\gamma_5$ and use the standard
property $\gamma_5^2=1$. So e.g. the correlator of two pseudoscalar
currents automatically coincides in the considered approach with
the correlator of two scalar currents.

{\bf 3. The singlet axial current.}
Let us consider now the case of the singlet axial current which
we define in the analogy with the non-singlet current (\ref{adc}):
\begin{equation}
\label{sa}
J^5_\mu = \overline\psi \gamma_\mu \gamma_5 \psi=
i\frac{1}{3!}\varepsilon_{\mu\nu_1\nu_2\nu_3}
\overline\psi \gamma_{\nu_1} \gamma_{\nu_2} \gamma_{\nu_3} \psi.
\end{equation}

It is known that the singlet axial current is nontrivially renormalized
because of the axial anomaly and the
renormalization constant of the singlet axial current
is nontrivial at the two-loop level \cite{adler},\cite{kod}.
To receive the correct renormalized
singlet axial current we need, as in the previous
cases, to introduce the finite renormalization constant $Z_5^s$ in addition to
the standard ultraviolet renormalization constant $Z_{MS}^s$ within
the MS-scheme:
\begin{eqnarray}
\label{rens}
(J^5_\mu)_R=Z_5^sZ_{MS}^s(J^5_\mu)_B=Z_5^sZ_{MS}^s
i\frac{1}{3!}\varepsilon_{\mu\nu_1\nu_2\nu_3}
\overline{\psi}_B \gamma_{\nu_1} \gamma_{\nu_2} \gamma_{\nu_3} \psi_B.
\end{eqnarray}
We can calculate within the MS-scheme the renormalization constant
$Z_{MS}^s$ of the singlet current at the three-loop level:
\begin{eqnarray}
\label{zas}
Z_{MS}^s  =  1
       + a^2[C_FC_A (\frac{22}{3\epsilon} )
       + C_Fn_f (\frac{5}{3\epsilon} )]
       + a^3[C_FC_An_f (  - \frac{22}{27\epsilon^2} + \frac{149}{81\epsilon} )
\nonumber \\
       + C_FC_A^2 ( -\frac{484}{27\epsilon^2} + \frac{3578}{81\epsilon} )
       + C_Fn_f^2 ( \frac{20}{27\epsilon^2} + \frac{26}{81\epsilon} )
       + C_F^2C_A (  - \frac{308}{9\epsilon} )
       + C_F^2n_f (  - \frac{22}{9\epsilon} )].
\end{eqnarray}
Now the problem is how to fix the finite renormalization constant
$Z_5^s$. As in the previous cases one should restore within dimensional
regularization the
standard properties of the singlet current which exist in such a basic
regularization procedure as the Pauli-Villars regularization.
We can not anymore impose for this purpose the coincidence of the axial
and vector vertices (\ref{recep}) because the singlet current generates closed
fermion loops
and we cannot anticommute $\gamma_5$ outside the singlet axial vertex.

To fix $Z_5^s$ one can require within dimensional regularization the
conservation of the one-loop character
\cite{ab} of the operator relation of the axial anomaly which is valid
in the Pauli-Villars regularization:
\begin{eqnarray}
\label{anom}
(\partial_\mu J^5_\mu)_R=a\frac{n_f}{2}(G\tilde{G})_R,
\end{eqnarray}
where
$G\tilde{G}=\varepsilon_{\mu\nu\lambda\rho}G_{\mu\nu}^aG_{\lambda\rho}^a$
and
$G_{\mu\nu}^a=\partial_\mu A_\nu^a-\partial_\nu A_\mu^a+g f^{abc}A_\mu^b
A_\nu^c$ is the gluonic field strength tensor.

Let us consider in detail renormalizations of both sides of the anomaly
relation.
The divergence $\partial_\mu J^5_\mu$ is renormalized multiplicatively in the
same way (\ref{rens}) as the current $J^5_\mu$ itself:
\begin{eqnarray}
\label{divr}
(\partial_\mu J^5_\mu)_R=Z_5^sZ^s_{MS}(\partial_\mu J^5_\mu)_B.
\end{eqnarray}
But the operator $G\tilde{G}$ mixes under renormalization:
\begin{eqnarray}
\label{rhsren}
(G\tilde{G})_R=Z_{G\tilde{G}}(G\tilde{G})_B+Z_{GJ}(\partial_\mu J^5_\mu)_B.
\end{eqnarray}
It is known \cite{et}\cite{bms} that to explain why
$\partial_\mu A_\mu$ does not mix
under renormalization and $G\tilde{G}$ does mix it is convenient to represent
$G\tilde{G}$ as the divergency of the axial gluon current:
\begin{eqnarray}
\label{k}
G\tilde{G} &=& \partial_\mu K_\mu,
\nonumber \\
K_\mu &=&4 \varepsilon_{\mu\nu_1\nu_2\nu_3}(A_{\nu_1}^a\partial_{\nu_2}
A_{\nu_3}^a+\frac{1}{3}gf^{abc}A_{\nu_1}^a
A_{\nu_2}^b A_{\nu_3}^c).
\end{eqnarray}
The current $K_\mu$ is not gauge-invariant. That is why the gauge invariant
current $J^5_\mu$ in some "good" gauges (e.g. axial gauge or
background-field
gauge) can not mix with the gauge-variant operator $K_\mu$. Then
the divergences $\partial_\mu J^5_\mu$ and $G\tilde{G}$ of the current
$J^5_\mu$ and
$K_\mu$ are known to be renormalized as the current themselves.
So $\partial_\mu A_\mu$ does not mix with $G\tilde{G}$ in these "good"
gauges either.
But since both divergences are gauge invariant this non-mixing is valid
in any gauge.

To understand the restrictions on the renormalization constants
$Z_{G\tilde{G}}$ and $Z_{GJ}$ it is
useful to take the renormalization group divergence of eq.(\ref{rhsren}):
\begin{eqnarray}
\mu^2\frac{d}{d\mu^2}(G\tilde{G})_R=
\gamma_{G\tilde{G}}(G\tilde{G})_R +\gamma_{GJ}(\partial_\mu J^5_\mu)_R,
\end{eqnarray}
where anomalous dimensions are defined for the case of operator
mixing as follows:
\begin{eqnarray}
\label{andim}
(O_i)_R=Z_{ij}(O_j)_B,~\gamma_{ij}=(\mu^2\frac{d}{d\mu^2}Z_{ik})(Z^{-1})_{kj}=
-a\frac{\partial z^{(1)}_{ij}}{\partial a},~
Z_{ij}=1+\sum_{n=1}^{\infty}\frac{z^{(n)}_{ij}(a)}{\epsilon^n}.
\end{eqnarray}
{}From the renormalization invariance of the anomaly:
\begin{eqnarray}
\mu^2\frac{d}{d\mu^2}(\partial_\mu J^5_\mu)_R=
\mu^2\frac{d}{d\mu^2}a\frac{n_f}{2}(G\tilde{G})_R,
\end{eqnarray}
one receives now:
\begin{eqnarray}
\label{genrestr}
\gamma_J^s(\partial_\mu J^5_\mu)_R =
a\frac{n_f}{2}[(\frac{\beta}{a}+\gamma_{G\tilde{G}})
(G\tilde{G})_R+\gamma_{GJ}(\partial_\mu J^5_\mu)_R] =
(\frac{\beta}{a}+\gamma_{G\tilde{G}}+a\frac{n_f}{2}\gamma_{GJ})
(\partial_\mu J^5_\mu)_R.
\end{eqnarray}
{}From this equation one can assume the restrictions on the anomalous
dimensions:
\begin{eqnarray}
\label{restr}
\gamma_{G\tilde{G}}=-\frac{\beta(a)}{a},~~~
\gamma_{GJ}=(a \frac{n_f}{2})^{-1}\gamma_J^s.
\end{eqnarray}
Strictly speaking eq.(\ref{genrestr}) itself permits a more general
solution than
the solution given in eq.(\ref{restr}). But the direct calculation at the
two-loop level (see below eq.(\ref{two})) supports eq.(\ref{restr}).

To calculate $Z_5^s$ one can calculate matrix elements of the l.h.s. and r.h.s.
of the anomaly operator equation (\ref{anom}) between gluon states
(so the famous anomalous triangle one-loop diagrams plus higher loops appear
in the l.h.s.). To be more precise
we calculate for the l.h.s of eq.(\ref{anom}) the following Green-function:
\[ <A ~\partial_\mu J^5_\mu~ A> =\]
\begin{eqnarray}
\label{lhs}
= R_{\overline{MS}}~~
\varepsilon_{\lambda\rho\nu\sigma}\frac{p_\nu}{p^2}
(\frac{\partial}{\partial q_\sigma})
\int d^4 xd^4 y  e^{ipx+iqy} < T{A_\lambda^a (x)
\partial_\mu J^5_\mu(y)
A_\rho^a(0) }>\mid_{q=0}^{amputated},
\end{eqnarray}
where 'amputated' means that
one-particle-irreducible diagrams with amputated external gluon legs
are considered. As it was explained above, the essential point of the
calculation is that
the product of two $\varepsilon$-tensors can be substituted by the determinant
of metric tensors which can be taken as $D$-dimensional ones inside
$R$-operation.
Some typical three-loop diagrams contributing to eq.(\ref{lhs})
are shown in fig.1.

The result of the three-loop calculation in the $\overline{MS}$-scheme is:
\begin{eqnarray}
\label{lhsres}
<A \partial_\mu J^5_\mu A> =
       24 n_f a \{ 1
       +a[ C_F( 4 )  + C_A ( 6 + 2\xi - \frac{1}{2}\xi^2 )]
       +a^2[ C_FC_A(\frac{323}{9} + 8\xi - 2\xi^2 )
\nonumber \\
       + C_Fn_f(  - \frac{349}{18} + 8\zeta_3 )
       + C_F^2(  - 6 )
+ C_An_f( - \frac{343}{24} - 12\zeta_3 + \frac{22}{9}\xi - \frac{5}{9}\xi^2 )
\nonumber \\
+ C_A^2( \frac{4537}{48} + 4\zeta_3  + 3\zeta_3\xi- \frac{1}{4}\zeta_3\xi^2
 + \frac{2467}{288}\xi
 - \frac{131}{72}\xi^2 - \frac{7}{8}\xi^3 + \frac{3}{16}\xi^4 )]\},
\end{eqnarray}
where $\xi$ is the gauge parameter in an arbitrary covariant gauge so
the gluon propagator is $(g_{\mu\nu}-\xi\frac{q_\mu q_\nu}{q^2})/q^2$.
We omitted in this result terms with $\log (p^2/\mu^2)$.

For the r.h.s of the anomaly equation (\ref{anom}) we calculate
at the two-loop level the analogous matrix element,
as for the l.h.s (\ref{lhs}). Some typical two-loop diagrams contributing to
the r.h.s can be obtained from the diagrams of fig.1 by shrinking the upper
fermion loop into a point.
Renormalization is done according to (\ref{rhsren}). The necessary
approximations for the renormalization constants are:
\begin{eqnarray}
\label{two}
Z_{G\tilde{G}}&=&1+a[\frac{1}{\epsilon}(- \frac{11}{3}C_A + \frac{2}{3}n_f)]
\nonumber \\ &&
+a^2[\frac{1}{\epsilon^2}(- \frac{44}{9}C_An_f + \frac{121}{9}C_A^2
 + \frac{4}{9}n_f^2)
+\frac{1}{\epsilon}
( C_Fn_f + \frac{5}{3}C_An_f - \frac{17}{3}C_A^2)],
\nonumber \\
Z_{GJ}&=&a\frac{1}{\epsilon}12C_F.
\end{eqnarray}
The validity of the restrictions of
eq.(\ref{restr}) is confirmed in this approximation. \\
The two-loop result in the $\overline{MS}$-scheme is:
\begin{eqnarray}
\label{rhsres}
\noindent
a \frac{n_f}{2}<A~ G\tilde{G}~ A> &=&
24 n_f a \{ 1
       + a [C_A ( 6 + 2\xi - \frac{1}{2}\xi^2 )]
       + a^2[ + C_Fn_f (  - \frac{53}{3} + 8\zeta_3 )
\nonumber \\ &&
+ C_An_f( - \frac{343}{24}- 12\zeta_3 + \frac{22}{9}\xi - \frac{5}{9}\xi^2  )
+C_A^2 ( \frac{4537}{48}+ 4\zeta_3 + 3\zeta_3 \xi
\nonumber \\ &&
 - \frac{1}{4}\zeta_3 \xi^2
+ \frac{2467}{288}\xi
 - \frac{131}{72}\xi^2 - \frac{7}{8}\xi^3 + \frac{3}{16}\xi^4  )]\}.
\end{eqnarray}
One can see that without the finite renormalization constant $Z_5^s$ the l.h.s
of the anomaly relation (\ref{lhsres}) and the r.h.s. (\ref{rhsres})
do not agree within the $\overline{MS}$-scheme.
We can obtain the desired finite constant $Z_5^s$
restoring the one-loop character of the anomaly relation, i.e.
dividing (\ref{rhsres}) by  (\ref{lhsres}):
\begin{eqnarray}
\label{z5s}
Z_5^s= 1
       + a  (  - 4C_F )
       + a^2 ( 22C_F^2 - \frac{107}{9}C_AC_F +\frac{31}{18}C_Fn_f ).
\end{eqnarray}
We should stress that the difference between the singlet constant $Z_5^s$ and
the non-singlet constant $Z_5^{ns}$ is only in the $C_Fn_f$-term.
This difference is due to only the light-by-light-scattering type diagrams
(the type shown first on fig.1).
The independence of the obtained finite constant $Z^s_5$ on $\log(p^2/\mu^2)$
gives strong confirmation that both sides of the anomaly relation are
really matched.

Thus both sides of the
axial anomaly relation receive the non-trivial higher order corrections
if one considers their matrix elements. But these corrections are matched and
the one-loop character of the operator equation (which is valid in the
Pauli Villars regularization) can be preserved in dimensional regularization.

Now we can calculate the full anomalous dimension of the singlet current
in $O(a^3)$ approximation:
\begin{eqnarray}
\label{gammaA}
\gamma_J^s(a) &=& \mu^2\frac{d\log (Z_5^sZ_{MS}^s)}{d\mu^2}
\nonumber \\
&=&
       + a^2  (  - 6C_Fn_f )
       + a^3  (  - \frac{142}{3}C_FC_An_f + \frac{4}{3}C_Fn_f^2 + 18C_F^2n_f ).
\end{eqnarray}
The first term agrees with the calculation \cite{kod}.

It is interesting to consider now the transition to the QED case in the
result (\ref{lhsres}) for the matrix element of the divergency
$\partial_\mu J^5_\mu$
between gluon states, multiplied finally by the finite constant
$Z_5^s$.
The transition
from the QCD case to the QED case can be done by simple substitutions:
$C_A=0,~ C_F=1,~ n_f/2=n_f,~\alpha_s=\alpha$.
Making this substitutions we find that the only surviving contributions to
the matrix element of $\partial_\mu J^5_\mu$ between two photons
are the famous one-loop triangle diagram and the three-loop diagrams of the
light-by-light-scattering type (the first type on fig.1).
The fact that these light-by-light-scattering
diagrams give a non-zero contribution to the matrix element
of $\partial_\mu J^5_\mu$
between two photon states was discovered
originally by direct calculation in \cite{aj}.
These three-loop diagrams give the correction to the width of the neutral
pion decay into two photons.

Thus in all calculations involving the singlet axial current one can apply the
calculational power of dimensional regularization by using
for the singlet axial current the prescription (\ref{rens})
with corresponding renormalization constants given in (\ref{zas}) and
(\ref{z5s}).

The generalization of the considered $\gamma_5$-prescription for the
massive case is straightforward. Within the MS-scheme the ultraviolet
renormalization constants do not depend on masses \cite{collins2}.
The same is valid for the finite renormalization constants $Z_5$, so the
obtained finite constants can be applied also in the massive case.
\[ \]
I am grateful to the collaborators of the Theory group of
NIKHEF-H for helpful discussions.

\end{document}